\documentclass[%
 reprint,
 amsmath,amssymb,
 aps,
]{revtex4-2}

\usepackage{graphicx}% Include figure files
\usepackage{dcolumn}% Align table columns on decimal point
\usepackage{bm}% bold math
\begin{document}

\preprint{APS/123-QED}

\title{Robust chaos in a totally symmetric network of four phase oscillators}% Force line breaks with \\

\author{Efrosiniia Karatetskaia$^1$}
\author{Alexey Kazakov$^1$}%
\author{Klim Safonov$^1$}
\author{Dmitry Turaev$^2$}

\affiliation{$^1$National Research University Higher School of Economics,
25/12 Bolshaya Pecherskaya Ulitsa, 603155 Nizhny Novgorod, Russia}%
\affiliation{$^2$Imperial College, London SW7 2AZ, United Kingdom}

\date{\today}% It is always \today, today,

\begin{abstract}
We provide conditions on the coupling function such that a system of 4 globally coupled identical oscillators has chaotic attractors, a pair of Lorenz attractors or a 4-winged analogue of the Lorenz attractor. The attractors emerge near the triple instability threshold of the splay-phase synchronization state of the oscillators. We provide theoretical arguments and verify numerically, based on the pseudohyperbolicity test, that the chaotic dynamics are robust with respect to small, e.g. time-dependent, perturbations of the system. The robust chaoticity should also be inherited by any network of weakly interacting  systems with such attractors.
\end{abstract}

\maketitle

\paragraph*{Introduction.}\label{sec:intro}

The study of emergent behavior in networks of coupled oscillators is an important problem with application to various fields of science and engineering, including neuroscience \cite{Izh07, ashwin2016mathematical, taylor2022noise}, molecular and cellular biology \cite{richard1996acetaldehyde, prindle2012sensing}, medical engineering \cite{peskin1975mathematical, garashchuk2020synchronous}, chemistry and material science \cite{K03, T09, toiya2010synchronization, yan2012linking}, optics \cite{kourtchatov1995theory, kozyreff2000global, takemura2020designs}, solid state physics \cite{benz1991coherent, wiesenfeld1998frequency, mukhopadhyay2023superconductivity}, mechanics \cite{martens2013chimera, belykh2017foot, belykh2021emergence}, etc.

The idea that a diffusive coupling in a system of identical damped oscillators can create instabilities leading to the symmetry breaking and the pattern formation can be traced back to Turing~\cite{T52}. The emergence of non-trivial, even chaotic, dynamics in such systems at an intermediate coupling strength is now a well-established fact~\cite{Sm76, DIR07, T23}. Without the damping (i.e., when the individual systems in the network have an asymptotically stable periodic orbit), non-trivial patterns of collective behavior of the oscillators phases are formed at an arbitrarily weak coupling \cite{ashwin1992dynamics, swift1992averaging} (the strong coupling leads to synchronization \cite{winfree1967biological, kuramoto1975self, afraimovich1986stochastic, berner2023synchronization}).

The dynamics of a totally symmetric
network of weakly coupled identical oscillators are modelled by the generalized Kuramoto model~\cite{kuramoto1975self, ashwin1992dynamics, swift1992averaging}
\begin{equation}
\dot{\phi}_j=\omega+\frac{1}{N}\, \sum_{i=1}^{N} g(\phi_i-\phi_j), \ \ j=1,\ldots, N,
\label{eq_Kuramoto}
\end{equation}
where $\phi_j\in S^1$ is the phase of the $j$-th oscillator, $N$ is the number of the oscillators, $\omega$ is the common frequency, and the $2\pi$-periodic function $g(\phi)$ describes the coupling.

In the classical Kuramoto-Sakaguchi model~\cite{kuramoto1975self,sakaguchi1986soluble}, one puts $g(\phi) =A \sin(\phi + \xi)$. This is an important special case, describing the interaction of small amplitude oscillations, e.g. when the parameters of an individual oscillator are close to the Andronov-Hopf bifurcation at which the small-amplitude stable periodic regime was created, cf.~\cite{ashwin2016hopf}. For such single-harmonic function $g$, regardless of the number of oscillators, the dynamics are either completely integrable (quasiperiodic, when $g(\phi)=A\cos(\phi)$ \cite{watanabe1993integrability}) or converge to a synchronized state \cite{Str00, watanabe1994constants}.

However, when the oscillators are far from the Andronov-Hopf bifurcation and the oscillations' amplitude is not small, the coupling function $g$ can be arbitrary. This strongly influences the network dynamics. Thus, for $N\geq 4$, adding the second Fourier harmonic to $g(\phi)$ can lead to apparently chaotic dynamics \cite{Ashwin2007, Bick11, Bick16, Grines23}.

Note that the previously observed chaotic attractors in the homogeneous network of oscillators \eqref{eq_Kuramoto} are not robust with respect to a variation of parameters. This is clearly demonstrated by the Lyapunov diagrams in Refs.~\cite{Bick11, grines2022origin, Grines23}, where parameter values corresponding to chaos (the positive top Lyapunov exponent) alternate with those corresponding to trivial attractors. Such non-robustness is a quite common feature of chaotic dynamics in general and is often observed in models of various nature \cite{AfrShil1983, GKT21}. The existence of the multitude of stability windows in the parameter space makes it impossible to be certain whether a regime observed in a numerical experiment is indeed chaotic, or it is just a long transient which could eventually degenerate into a regular regime such as a stable periodic orbit or a stable stationary state.

In this Letter, we show that system \eqref{eq_Kuramoto} with $N=4$ oscillators can demonstrate, in certain regions of parameter values, a \textit{robust chaotic behavior}, with no stability windows. We give a rigorous proof of the existence of these regions and provide explicit conditions on the coupling function $g(\phi)$ which allow one to find them, see conditions \eqref{triple_zero_condition}, \eqref{attractor_condition}.
We apply the theoretical results to an example of the coupling function with 4 Fourier modes:
$g(\phi)=\sum_{k=1}^4 A_k\sin (k\, \phi+\xi_k)$,
for which we determine the regions of robust and non-robust chaos numerically (we also show that the presence of the 4th harmonic is necessary for the robust chaos to appear).

The robust chaotic attractors we find are 4-winged analogues of the classical Lorenz attractor, see Refs.~\cite{KKST24a, KKST24b}. Importantly, the robustness of chaos means here more than mere stability with respect to small variations of parameters. These attractors remain chaotic under time-dependent perturbations (periodic, quasiperiodic), and any network of weakly coupled, identical systems with attractors of this type is also robustly chaotic.

\paragraph*{Pseudohyperbolicity.}

To establish the robustness of chaos, we use the notion of pseudohyperbolicity introduced in \cite{TS98}. It is based on the computation of Lyapunov exponents and the verification of the continuity of certain fields of invariant Lyapunov subspaces \cite{kuptsov2012theory, ginelli2013covariant, pikovsky2016lyapunov} at points of the attractor.

Given an orbit in a chaotic attractor, one computes its Lyapunov exponents $\lambda_1, \dots, \lambda_N$ (indexed in the decreasing order).
Let $\lambda_1\geq \dots \geq \lambda_p \; > \;\lambda_{p+1} \geq \dots \geq \lambda_N$ for some $p$. Then, at each point of the orbit there are linear subspaces $E_1$ and $E_2$ (${\rm dim}(E_1)=p$, ${\rm dim}(E_2)=N-p$) such that the field of these subspaces is invariant with respect to the linearization of the system along the trajectory. For the linearized system, the exponential growth (or contraction) rate for the vectors in $E_1$ is bounded from below by $\lambda_p$ and it is bounded from above by $\lambda_{p+1}$ for the vectors in $E_2$. The foundational Oseledets theorem \cite{oseledets1968multiplicative} guarantees the existence of Lyapunov exponents and the corresponding invariant subspaces $E_{1,2}$ for a representative orbit in the attractor, but it does not distinguish between the following two possibilities: the fields of subspaces $E_{1,2}$ either admit an extension by continuity to the closure of the orbit, or not. The general theory only gives a measurable dependence of $E_{1,2}$ on the point in the attractor, and not continuous a priori.

When the dependence is {\em continuous}, the angle between $E_1$ and $E_2$ stays bounded away from zero and this splitting into a direct sum of two invariant subspaces persists at all small perturbations \cite{bonatti2004dynamics}. In this case, the attractor is called pseudohyperbolic if $\lambda_1+\dots+\lambda_p > 0$, i.e., {\em the total volume is expanded in} $E_1$. This property also survives small perturbations. Since the volume expansion in an invariant subspace automatically guarantees the positivity of the top Lyapunov exponent for every orbit in the attractor, the pseudohyperbolicity implies the robustness of chaotic dynamics \cite{TS08}.

All known examples of robustly chaotic attractors are pseudohyperbolic. This includes hyperbolic and partially-hyperbolic attractors \cite{anosov1995dynamical, pesin1982gibbs, kuznetsov2005example}, Lorenz-like attractors \cite{lorenz1963deterministic, guckenheimer1979structural, ABS77, ABS82, Tuc99}, and others \cite{TS98, GKT21, GGKS21, GKKK22, barros2024upper, kazakov2024numerical}, as well as their discretized versions and time-periodic perturbations \cite{GOST05, TS08}. On the other hand, the theory of homoclinic tangencies suggests that stability windows can emerge easily when chaotic attractors without a pseudohyperbolic
structure are perturbed \cite{newhouse1974diffeomorphisms, AfrShil1983, gonchenko2008dynamical}. Therefore, we proposed in Ref.~\cite{GKT21} the pseudohyperbolicity as a universal criterion for the robust chaoticity.

\paragraph*{Numerics.}
The robustness allows one to be sure that a numerically produced ``strange attractor'' corresponds indeed
to a truly chaotic regime and is not a transient or an effect of the inevitable noise. The pseudohyperbolicity criterion makes the numerical verification of the robustness quite straightforward. First, one computes the Lyapunov exponents, chooses $p$ such that $\lambda_p >\lambda_{p+1}$ and $\lambda_1+\dots+\lambda_p >0$ (the volume expansion in $E_1$). Then, the rest is to check the continuous dependence of the corresponding subspaces $E_{1,2}$ on a point in the attractor. The continuity can be verified directly \cite{GKT21}, or by evaluating the angles between $E_1$ and $E_2$ and checking that they are bounded away from zero \cite{kuptsov2012fast, kuptsov2018lyapunov, GKKT21}.

\begin{figure}[h]
\begin{minipage}[h]{1\linewidth}
\center{\includegraphics[width=1\linewidth]{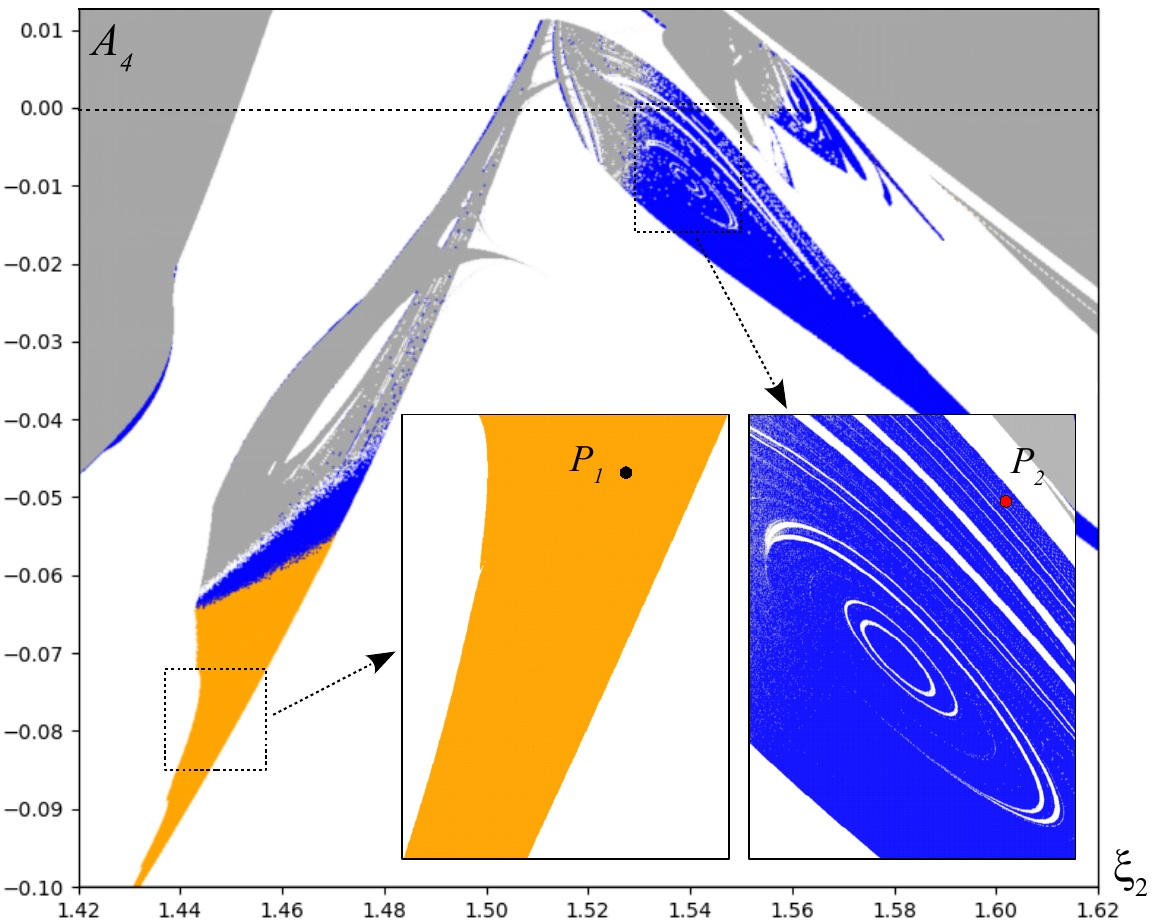}}
\end{minipage}
\caption{\footnotesize Lyapunov diagram for system \eqref{eq_theta_system} in the parameter plane $(\xi_2, A_4)$ for $\xi_1 = -0.01, A_1 = -1$. At each parameter value from the grid $1000 \times 1000$, we compute the Lyapunov exponents $\lambda_1\geq \lambda_2 \geq \lambda_3$. If $\lambda_1>0$, we additionally check that $\lambda_1+\lambda_2 >0$, $\lambda_2>\lambda_3$, and the minimal angle $\alpha$ between the Lyapunov subspaces $E_1$ and $E_2$ is bounded away from zero. In the orange region these conditions are fulfilled, i.e., we have a robust chaotic attractor. In the blue region, $\alpha \approx 0$, so the pseudohyperbolicity conditions are violated, and the chaos is not robust; one can see numerous stability windows in the blue region. In the white and gray regions, the attractor is a periodic orbit or a stationary state, respectively.}
\label{fig1}
\end{figure}

We applied this strategy to system \eqref{eq_Kuramoto} with
\begin{equation}
g(\phi)=A_1\sin(\phi+\xi_1)+\sin(2\phi+\xi_2) + A_4\sin(4\phi).
\label{eq_gFunc}
\end{equation}
By the phase-shift symmetry, it is natural to introduce phase differences
$\theta_j=\phi_j-\phi_4$ (we let $\theta_4=0$ for the convenience of notation). System \eqref{eq_Kuramoto} becomes 3-dimensional:
\begin{equation}
\dot{\theta}_j=\frac{1}{4}\, \sum_{i=1}^{4} \bigl[g(\theta_i-\theta_j)- g(\theta_i)\bigr], \ \ j=1,2,3.
\label{eq_theta_system}
\end{equation}
We put $A_1=-1$, $\xi_1 = -0.01$ and did the following experiment: at each value of the parameters $(\xi_2,A_4)\in [1.62,..,1.92]\times [-0.1,...,0.013]$ from the grid of $1000 \times 1000$ points, we took the orbit with the initial condition $(\theta_1, \theta_2, \theta_3) = (1, 1.5, 3)$ and consider the segment of this orbit with $t\in [10^4,10^5]$ as an approximation of the attractor of system \eqref{eq_theta_system}. We computed the Lyapunov exponents $\lambda_1\geq \lambda_2\geq \lambda_3$,
and marked the attractor as chaotic if $\lambda_1\geq 0.005 >0$ (blue points
in Fig.~\ref{fig1}). We consider the attractor robustly chaotic (orange in Fig.~\ref{fig1}), if the orbit passes the pseudohyperbolicity test. For that, we checked whether $\lambda_1+\lambda_2\geq 0.005$, $\lambda_2  >  \lambda_3$, and the angle between the corresponding two-dimensional volume-expanding Lyapunov subspace $E_1$ and one-dimensional contracting Lyapunov subspace $E_2$ is bounded away from zero: $\alpha \geq 0.005$ at each point of the numerically generated attractor.

The resulting diagram is shown in Fig.~\ref{fig1}. In the orange region, the pseudohyperbolicity conditions are satisfied. There are no stability windows, as is clearly visible in the left inset. In the blue regions, we have both the area-expansion $\lambda_1+\lambda_2>0$ and dominated splitting conditions $\lambda_2 > \lambda_3$, but the angle $\alpha$ is not separated from zero. Therefore, chaotic attractors in these regions are not pseudohyperbolic. This is in agreement with the presence of stability windows (white color -- stable periodic orbits; gray -- a stable stationary state) inside the regions of chaoticity, see the right inset in Fig.~\ref{fig1}.

\begin{figure}[ht]
\begin{minipage}[h]{1\linewidth}
\center{\includegraphics[width=1\linewidth]{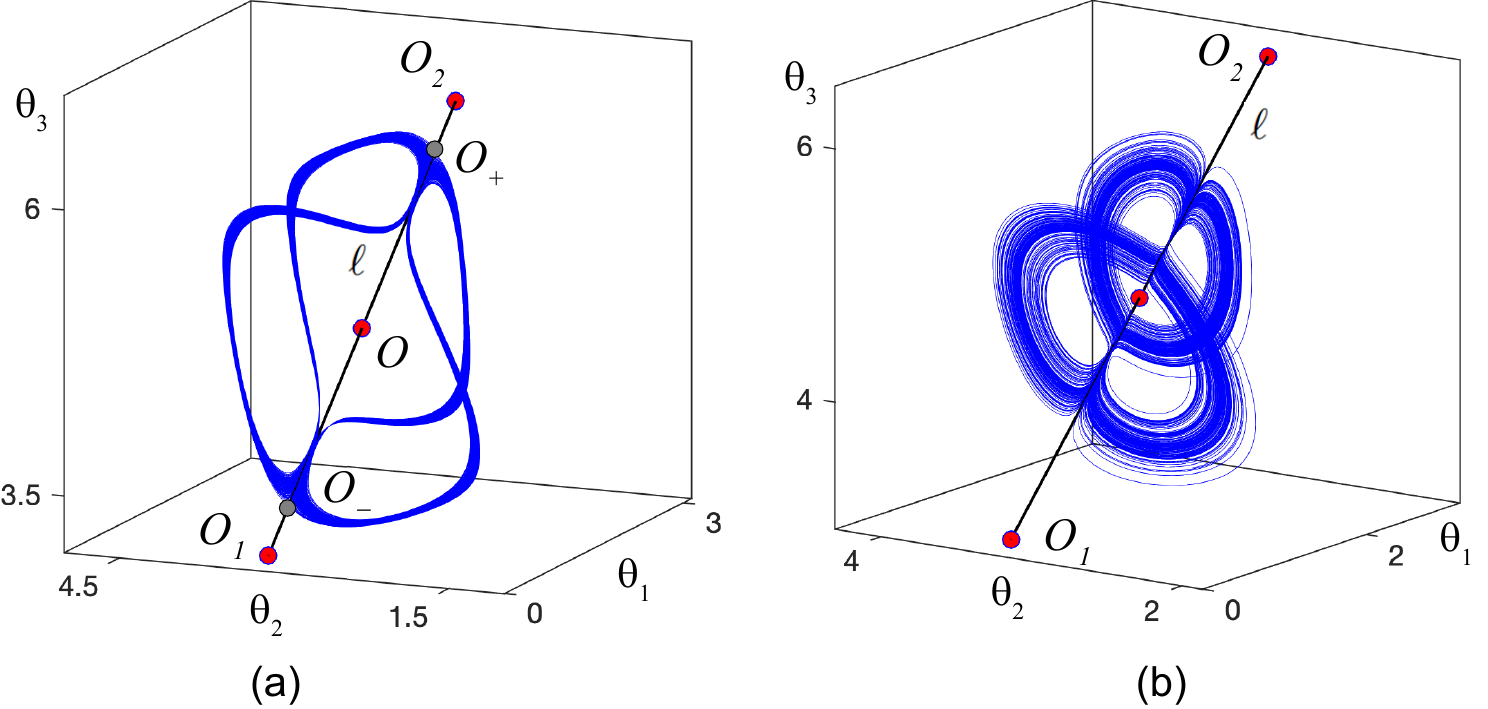}}
\end{minipage}
\caption{\footnotesize (a) Pseudohyperbolic attractor at $(\xi_2, A_4) = (1.45, -0.074)$ (the point $P_1$ taken in the orange region in  Fig.~\ref{fig1}); (b) non-pseudohyperbolic attractor at $(\xi_2, A_4) = (1.545, -0.003)$ (the point $P_2$ taken in the blue region in  Fig.~\ref{fig1}). At Fig.~b, the invariant line $l$ contains three equilibria: stable equilibria $O_{1,2}$ and the equilibrium $O$, which is unstable in $l$ and is stable in a transverse direction. At Fig.~a, the equilibria $O_{1,2}$ are unstable in $l$, and the invariant line contain two additional equilibria $O_{\pm}$, which are stable in $l$ and unstable in a transverse direction.}
\label{fig2}
\end{figure}

The chaotic attractors from the orange and blue regions (at points $P_1$ and $P_2$, respectively) are shown in Fig.~\ref{fig2}. Note that they have quite similar shape, but the attractor in Fig.~\ref{fig2}a is robustly chaotic, whereas the attractor in Fig.~\ref{fig2}b fails the pseudohyperbolicity test. Hence, one cannot reliably conclude whether this is a genuine chaos or a chaotic transient sustained by the numerical noise.
Note that the line $A_4=0$ (that corresponds to the only first 2 harmonics in the coupling function $g(\phi)$) does not intersect the orange region of robust chaos. Thus, adding the 4th harmonic is crucial for the creation of verifiable chaotic attractors in system \eqref{eq_Kuramoto}.

\paragraph*{Four-winged attractors.} The chaotic attractors, which we found numerically for a particular choice of the coupling function, emerge under fairly general conditions. They are examples of 4-winged Lorenz-like attractors (``Simo angels''), a geometric model for which is described in Refs.~\cite{KKST24a, KKST24b}. Their characteristic shape is due to the $\mathbb{Z}^4$-symmetry in the system, permuting the attractor's wings.

The symmetry in system \eqref{eq_Kuramoto} with $N=4$ is generated by the permutation $S: (\phi_1,\phi_2,\phi_3,\phi_4) \mapsto
(\phi_4,\phi_1,\phi_2,\phi_3-2\pi)$; obviously, $S^4=id$. Note that, up to reordering the phases $\phi_j$, it is enough to consider initial conditions in the symplex $C: \{\phi_4\leq \phi_1\leq \phi_2\leq \phi_3\leq \phi_4 + 2\pi\}$. Since the boundaries $\phi_j=\phi_k \mod 2\pi$ are invariant with respect to system \eqref{eq_Kuramoto}, the orbits starting in $C$ can never leave it. The permutation $S$ also takes $C$ into itself. In system \eqref{eq_theta_system} for the phase differences $\theta$, the invariant symplex $C$ corresponds to $\hat C: \{0\leq \theta_1 \leq \theta_2 \leq \theta_3\leq 2\pi\}$; the permutation $S$ acts on $\hat C$ as
$(\theta_1,\theta_2,\theta_3)\mapsto (\theta_2-\theta_1,\theta_3-\theta_1,2\pi-\theta_1)$.

The line $\ell:\{\theta_3=\theta_1+\pi, \ \theta_2=\pi\}$ is $S$-invariant and contains the $S$-symmetric splay-phase state $O=(\pi/2,\pi,3\pi/2)$. The symmetry axis $\ell$ intersects the boundary of $\hat{C}$ at the stationary points $O_1=(0,\pi,\pi)$ and $O_2=(\pi,\pi,2\pi)$. If $O$, $O_1$, and $O_2$ are unstable in $\ell$, then the invariant line $\ell$ must have an $S$-symmetric  pair of equilibria $O_\pm$ which are stable in $\ell$, see Fig~\ref{fig2}a. The 4-winged attractors may exist when  $O_\pm$ are unstable in a direction transverse to $\ell$. In this case, the attractor can be defined as the prolongation of $O_\pm$ -- the set of all points attainable from saddle stationary points $O_+$ or $O_-$ by $\varepsilon$-orbits for all small $\varepsilon>0$ \cite{ruelle1981small,TS98,GT17}. From the computational perspective, this is the closure of numerically generated orbits that start close enough to $O_\pm$.

We say that the attractor has 4 wings, if the orbits from a small neighborhood of $O_+$ come to a small neighborhood of $O_-$ following one of the two unstable separatrices of $O_+$. By the symmetry, the orbits
from a neighborhood of $O_-$ return to a small neighborhood of $O_+$. Thus, the attractor has an absorbing neighborhood which consists of 2 balls around $O_\pm$ and 4 handles around the separatrices. The chaoticity of the attractor expresses itself in the randomness of the itinerary with which the orbits follow the handles. This is very similar to the standard 2-winged Lorenz attractors \cite{lorenz1963deterministic}, with the difference that the latter have 1, not 2, stationary points and the absorbing domain has 2, not 4, handles, see Fig.~\ref{fig4}.

\paragraph*{Triple instability and robust chaos.}
In papers \cite{KKST24a, KKST24b}, we studied bifurcations of triply unstable (3 zero eigenvalues) equilibrium states in $\mathbb{Z}^4$-symmetric systems and provided general conditions for the emergence of pseudohyperbolic 2-winged and 4-winged Lorenz-like attractors. We apply this theory to the bifurcations of the splay-phase state $O$.
The linearization matrix of system \eqref{eq_theta_system} at $O$ is
$$
\frac{1}{4}\, \cdot \begin{pmatrix}
-\sum_{k=1}^3 c_k &  c_1- c_2 &  c_2- c_3\\
c_3- c_1 & -\sum_{k=1}^3 c_k & c_1- c_3\\
c_2- c_1 & c_3- c_2 & -\sum_{k=1}^3 c_k
\end{pmatrix},
$$
where $c_k=g'\left( \pi k/2\right)$. The eigenvalues
$
\mu=-(c_1+c_3)/2$,
$-\gamma+i \beta =-(c_1+2c_2+c_3)/4\pm i\, (c_1-c_3)/4 \ \
$
vanish if
\begin{equation}\label{triple_zero_condition}
g'\left(\frac{\pi}{2}\right)=g'\left(\pi\right)=g'\left(\frac{3\pi}{2}\right)=0.
\end{equation}
Choose the coordinates \quad
$
u=(\theta_1-\theta_3+\pi)- i(\theta_2-\pi)$, \\ $z=\theta_1-\theta_2+\theta_3-\pi.
$
System \eqref{eq_theta_system} takes the form
\begin{equation*}\label{eq:FlowNormalFormGeneral}
\begin{aligned}
\dot{u}&=(-\gamma+i \beta )\, u+a_0zu^*  + a_1 u^2u^*+a_2 z^2 u + a_3 (u^*)^3+O_4, \\
\dot{z}&=\mu\, z+b_0\, u^2+b_0^*\, (u^*)^2 +b_1\, z^3+ b_2\, zuu^*+O_4,
\end{aligned}
\end{equation*}
where \quad $a_0=\frac{g''(\pi/2)+g''(3\pi/2)}{8}+i\, \frac{g''(\pi/2)-g''(3\pi/2)}{8},$\\
$
b_0=\frac{g''(\pi)}{8}-i\, \frac{g''(\pi/2)-g''(3\pi/2)}{16}$$, \
b_1=-\frac{g'''(\pi/2)+g'''(3\pi/2)}{48}.$

\vskip 0.2cm

By Theorems 1.1-1.3 from \cite{KKST24b}, when the conditions $b_1<0$, ${\rm Re}(a_0b_0)<0$, and ${\rm Im}(a_0b_0)\neq 0$ are fulfilled, the pseudohyperbolic 2-winged and 4-winged Lorenz-like attractors exist for open regions in the $(\gamma,\beta)$-plane for small $\mu>0$. This result implies that\\ {\it the generalized Kuramoto system with $N=4$ has pseudohyperbolic attractors for small $g'(\pi/2)+g'(3\pi/2)<0$, provided the following conditions hold}
\begin{align}\label{attractor_condition}
&g'''\left(\frac{\pi}{2}\right)+g'''\left(\frac{3\pi}{2}\right)>0, \nonumber \\
&\left[g''\left(\frac{\pi}{2}\right)-g''\left(\frac{3\pi}{2}\right)\right]^2+g''\left(\pi\right)\left[g''\left(\frac{\pi}{2}\right)+g''\left(\frac{3\pi}{2}\right)\right]<0, \nonumber \\
&g''\left(\frac{\pi}{2}\right)-g''\left(\frac{3\pi}{2}\right)\neq 0\\
&g''\left(\frac{\pi}{2}\right)-2g''\left(\pi\right)+g''\left(\frac{3\pi}{2}\right)\neq 0. \nonumber
\end{align}

Note that the function $g$ with only 3 first harmonics ($A_4=0$) cannot simultaneously satisfy conditions \eqref{triple_zero_condition}, \eqref{attractor_condition}. Moreover, if $A_4=0$, then the equilibria $O_{\pm}$ do not exist, see Fig.~\ref{fig2}b, and hence there are no Lorenz-like attractors.

\begin{figure}[h]
\begin{minipage}[h]{1\linewidth}
\center{\includegraphics[width=1\linewidth]{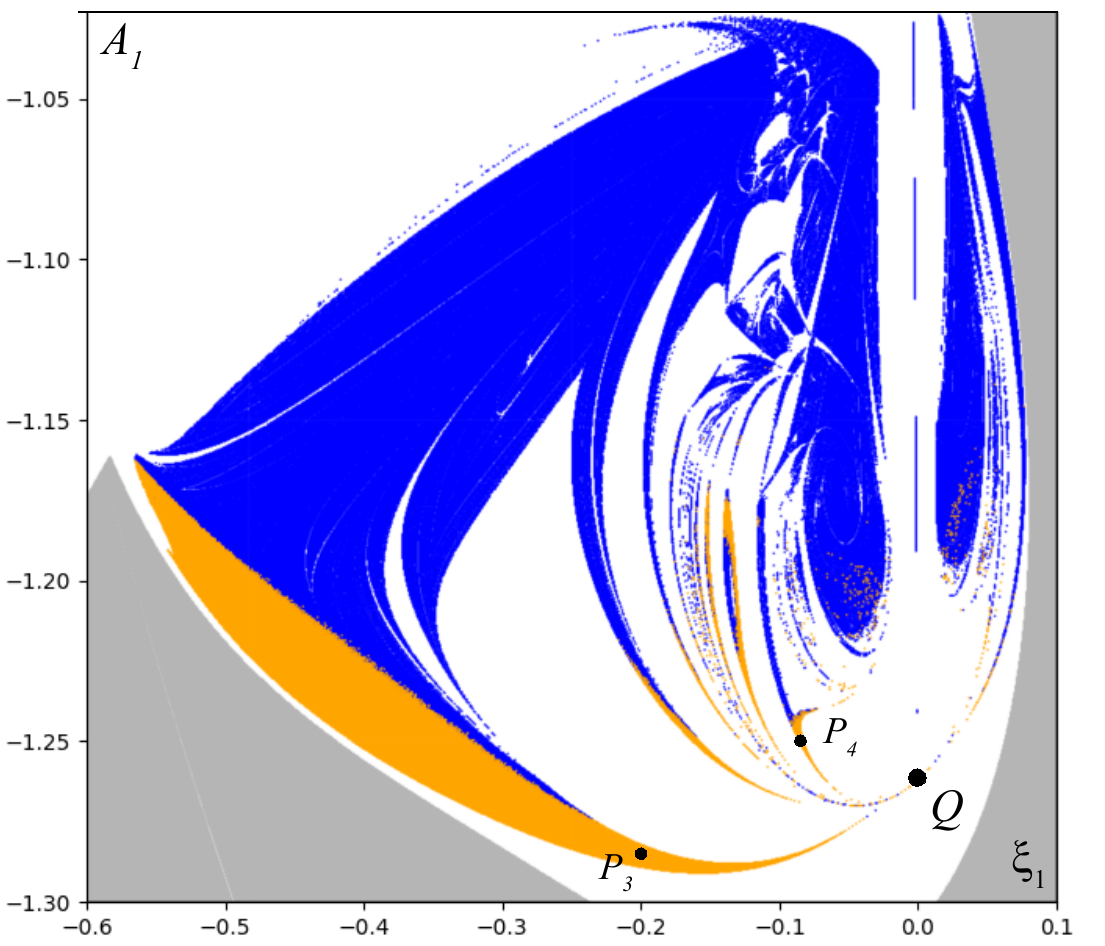}}
\end{minipage}
\caption{\footnotesize Lyapunov diagram for system \eqref{eq_theta_system} in the parameter plane $(\xi_1, A_1)$ for $\xi_2 = 1.81, A_4 = -0.125$. Regions of the existence of pseudohyperbolic attractors (orange) adjoin to the point $Q:(\xi_1,A_1)\approx (0,-1.27)$. The region containing the point $P_3$ corresponds to an $S$-symmetric pair of $2$-winged Lorenz attractors, see Fig.~\ref{fig4}a, and the region containing the point $P_4$ corresponds to a $4$-winged attractor, see Fig.~\ref{fig4}b.}
\label{fig3}
\end{figure}

On the contrary, in the presence of the 4th harmonic, conditions of the theorem are easily satisfied. For instance, for the coupling function $g$ given by formula \eqref{eq_gFunc} with $A_4=-0.125$, the splay-phase state $O$ has three eigenvalues at
$A_1=-1$, $\xi_1=0$, $\xi_2=\arccos\left(-0.25\right)\approx 1.823$, and inequalities \eqref{attractor_condition} are fulfilled. Therefore, system \eqref{eq_theta_system} has robustly chaotic attractors for $\xi_2$ close to $\arccos(-0.25)$ from below.

In Fig.~\ref{fig3}, we show the regions of the existence of pseudohyperbolic attractors in the $(\xi_1,A_1)$-plane for $\xi_2=1.81$, $A_4=-0.125$. To plot this diagram, we repeated the same experiment as was done for Fig.~\ref{fig1}, but with the initial conditions $(\theta_1,\theta_2,\theta_3)=(1.57,3.14,4.71)$ (near the splay-phase state $O$), and the tolerance bounds equal to $0.001$ for the Lyapunov exponents and $0.03$ for the angle $\alpha$ between the Lyapunov subspaces. The diagram is consistent with that for the theoretical normal form, see Fig.~6 in Ref.~\cite{KKST24a}. In particular, there are several regions corresponding to the 2-winged and 4-winged pseudohyperbolic attractors, see Fig.~\ref{fig4}. In agreement with Theorem 1.3 of Ref.~\cite{KKST24b}, these regions adjoin to the same point $Q:(\xi_1,A_1) \approx (0,-1.2625)$.

\begin{figure}[ht]
\begin{minipage}[h]{1\linewidth}
\center{\includegraphics[width=1\linewidth]{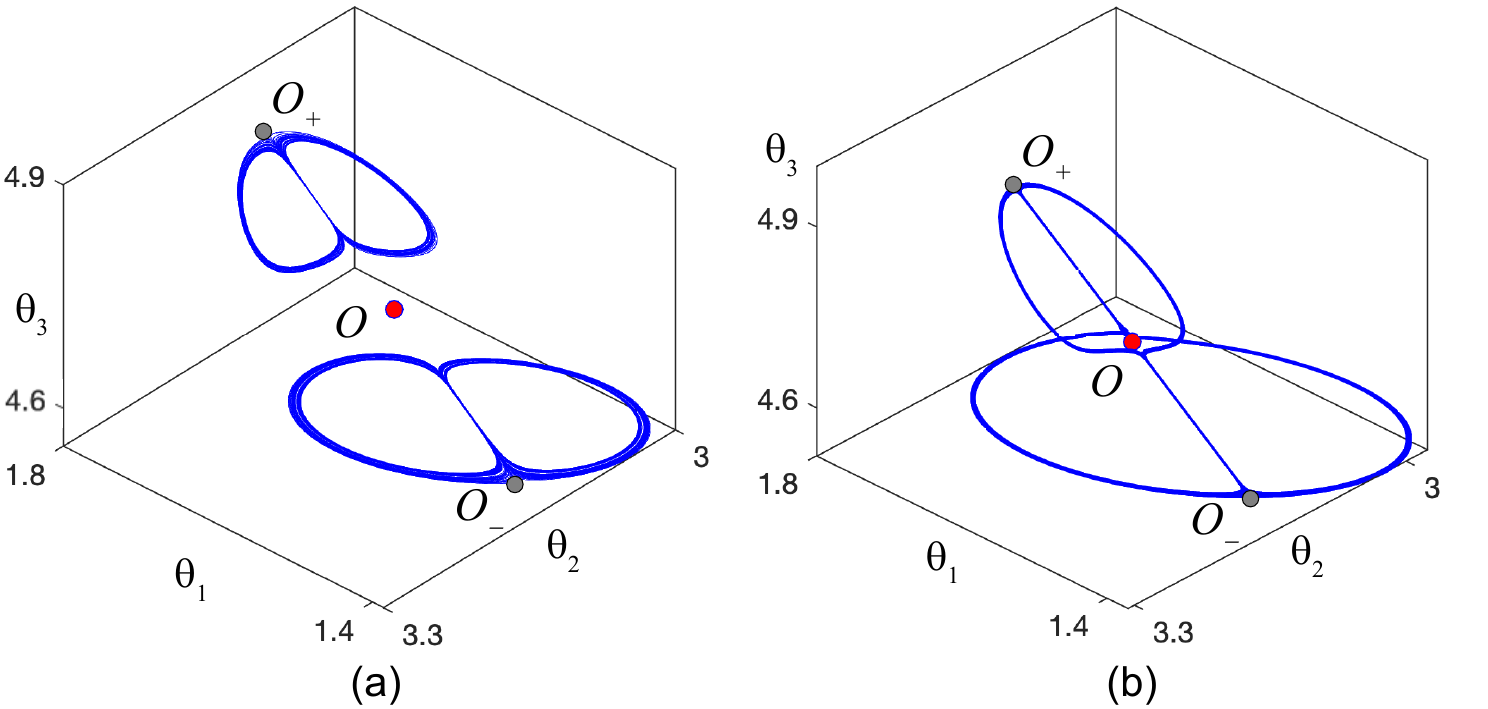}}
\end{minipage}
\caption{\footnotesize (a) a pair of 2-winged Lorenz attractors at $P_3:(\xi_1, A_1) = (-0.2,-1.285)$; (b) a 4-winged Lorenz-like attractor at  $P_4:(\xi_1, A_1) = (-0.085,-1.25)$.}
\label{fig4}
\end{figure}

\paragraph*{Conclusion.} There are two types of chaotic attractors: for pseudohyperbolic attractors
every orbit has positive top Lyapunov exponent and this property is robust with respect to small perturbations, and if the pseudohyperbolicity is violated, then only a typical orbit has positive top Lyapunov exponent and small perturbations can destroy chaos and create stable periodic orbits. Numerically, the pseudohyperbolicity (hence -- robust chaos)  can be reliably verified by augmenting the computation of Lyapunov exponents with the evaluation of angles between Lyapunov subspaces. We gave
rigorous analytic conditions for the emergence of pseudohyperbolic attractors in a system of
4 globally coupled identical phase oscillators due to a triple instability bifurcation of the splay-phase
synchronization state. The existence, in the space of parameters of the system, of open regions of robustly chaotic behavior was confirmed numerically.

The robustness implies that the found chaotic attractors survive any small changes in the coupling function, as well as the frequency detuning, breaking the symmetry of the coupling, etc. An important general fact is that weakly coupled identical systems with pseudohyperbolic attractors also have pseudohyperbolic attractors. Therefore, our results imply that any network of 4-oscillator systems of the type studied here also has pseudohyperbolic attractors, provided the interaction between the 4-oscillator clusters is weak enough. In this way, one can create large ensembles of weakly-interacting identical oscillators that demonstrate robustly chaotic behavior.

%\begin{acknowledgments}
\paragraph*{Acknowledgments.} This work was prepared within the framework of the Basic Research Program at HSE University and Leverhulme Trust grant RPG-2021-072.
%\end{acknowledgments}

%\bibliography{bibliography}% Produces the bibliography via BibTeX.
\providecommand{\noopsort}[1]{}\providecommand{\singleletter}[1]{#1}%

\end{document}